\documentclass[10pt,twocolumn,letterpaper]{article}

\usepackage{cvpr}              

\usepackage{multirow} 
\usepackage{booktabs} 
\usepackage{amsmath, amsfonts} 
\usepackage{bm} 
\usepackage{graphicx} 
\usepackage{wasysym} 
\usepackage{diagbox} 
\usepackage[linesnumbered,ruled,vlined]{algorithm2e}

\usepackage[dvipsnames]{xcolor}

\newcommand{\tul}[1]{\textit{\underline{#1}}}

\newcommand{\Tref}[1]{Table~\ref{#1}}
\newcommand{\Eref}[1]{Eq.~(\ref{#1})}
\newcommand{\Fref}[1]{Fig.~\ref{#1}}

\newcommand{\Aref}[1]{Alg.~\ref{#1}}

\definecolor{cvprblue}{rgb}{0.21,0.49,0.74}
\usepackage[pagebackref,breaklinks,colorlinks,citecolor=cvprblue]{hyperref}

\def\paperID{*****} 
\def\confName{CVPR}
\def\confYear{2024}

\title{Segue: Side-information Guided Generative Unlearnable Examples for Facial Privacy Protection in Real World}

\author{
    Zhiling Zhang$^{1}$, Jie Zhang$^{2}\dagger$, Kui Zhang$^{1}$, Wenbo Zhou$^{1}\dagger$, Weiming Zhang$^{1}$, Nenghai Yu$^{1}$\\
    $^{1}$University of Science and Technology of China  \\
    $^{2}$Nanyang Technological University\\
    {\tt\small\{zhilingzhang@mail., zk19@mail., welbeckz@, zhangwm@, ynh@\}ustc.edu.cn } \\
    {\tt\small jie\_zhang@ntu.edu.sg }\\
}
 
\begin{document}
\maketitle
\begin{abstract}
    The widespread use of face recognition technology has given rise to privacy concerns, as many individuals are worried about the collection and utilization of their facial data.
    To address these concerns, researchers are actively exploring the concept of ``unlearnable examples", by adding imperceptible perturbation to data in the model training stage, which aims to prevent the model from learning discriminate features of the target face. However, current methods are inefficient and cannot guarantee transferability and robustness at the same time, causing impracticality in the real world. 
    To remedy it, we propose a novel method called \textbf{Segue}: \textbf{S}id\textbf{e}-information guided \textbf{g}enerative \textbf{u}nlearnable \textbf{e}xamples. Specifically, we leverage a once-trained multiple-used model to generate the desired perturbation rather than the time-consuming gradient-based method. To improve transferability, we introduce side information such as true labels and pseudo labels, which are inherently consistent across different scenarios. For robustness enhancement, a distortion layer is integrated into the training pipeline. 
    Extensive experiments demonstrate that the proposed \textbf{Segue} is much faster than previous methods (1000$\times$) and achieves transferable effectiveness across different datasets and model architectures. Furthermore, it can resist JPEG compression, adversarial training, and some standard data augmentations. 
\end{abstract}    
\section{Introduction}
\label{sec: Introduction}
    Due to the rise of social media platforms like Twitter and Facebook, there has been a noticeable increase in the amount of facial data shared publicly, for fun or commercial purposes.
    Every coin has two sides. It becomes convenient for the unauthorized collection of individual facial data, which is a violation of public privacy \cite{Prabhu2020LargeID, ClearviewAI}.
    In addition, such facial data can be used to train various face analysis models such as face recognition models \cite{Jiang2016Face1, Ranjan2016face2, Yi2014webface, Liu2014CelebA}, which poses a threat to security-critical applications like authentication system \cite{Yin2021AdvMakeup, Yang2023adv3d}. Therefore, it is crucial to safeguard individual faces from unauthorized exploration. 
    
\begin{figure}[t]
    \centering
    \includegraphics[width=0.47\textwidth]{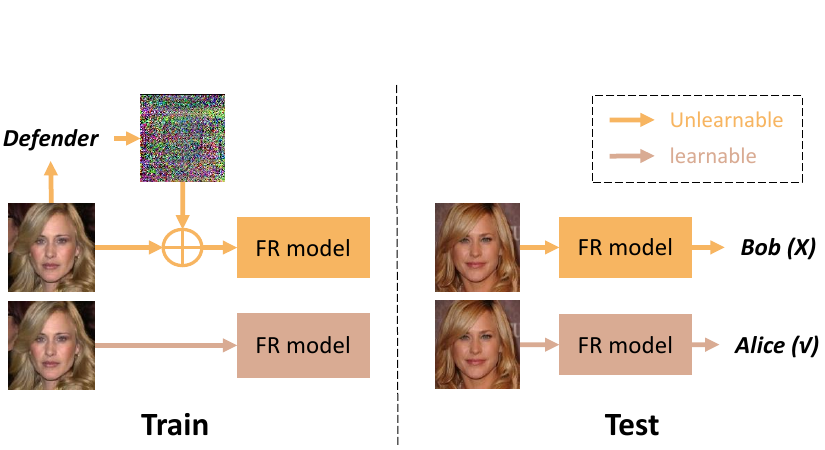}
    \caption{Illustration on leveraging the unlearnable examples for facial privacy protection.}
    \label{fig: scene}
    \vspace{-1em}
    \end{figure}

    Recent works focus on utilizing \textit{unlearnable examples} \cite{huang2021unlearnable, fu2022robust, ren2023transferable} to prevent attackers from training recognition models.  
    As shown in \Fref{fig: scene}, the defender adds some perturbations to the pristine image before releasing it, wherein the perturbed image is dubbed an unlearnable example. The attacker can only use the released unlearnable examples of Alice to train a facial recognition (FR) model, which fails in the inference stage (\ie, recognizing Alice as Bob). 
    The explanation for such a technique is that Neural networks are more inclined to learn shortcuts as discriminate features on classification tasks \cite{Geirhos2020ShortcutLI}, and the perturbation can be seen as a kind of shortcut.
    In a nutshell, we can leverage the unlearnable example for facial privacy protection.
    

    \begin{table*}[t]
    \setlength{\tabcolsep}{5.7mm}{
    \begin{tabular}{cccccc}
        \toprule Methods&Effectiveness&Imperceptibility&Transferability&Robustness&Efficiency\\
        \hline
        UE \cite{huang2021unlearnable}  &\CIRCLE&\CIRCLE&\RIGHTcircle&\RIGHTcircle&\Circle\\ 
        LSP \cite{Yu2021AvailabilityAC}&\RIGHTcircle&\RIGHTcircle&\RIGHTcircle&\RIGHTcircle&\CIRCLE \\
        RUE \cite{fu2022robust} &\CIRCLE&\CIRCLE&\RIGHTcircle&\CIRCLE&\Circle\\ 
        TUE \cite{ren2023transferable} &\RIGHTcircle&\CIRCLE&\RIGHTcircle&\RIGHTcircle&\Circle\\ 
        Ours &\CIRCLE&\CIRCLE&\CIRCLE&\CIRCLE&\CIRCLE\\
        \bottomrule
    \end{tabular}
    }
    \caption{We compare the proposed method with previous methods of unlearnable examples based on five requirements. We use \Circle, \RIGHTcircle, or \CIRCLE ~to indicate whether the method has no, partial, or full ability for each requirement.} 
    \label{tab: Comparison}
    \end{table*} 

    To generate unlearnable facial examples that can be applied in the real world, there are \textbf{five} requirements:
    1) \tul{Effectiveness}: the generated unlearnable examples shall make the FR model cannot recognize the corresponding clean examples, with accuracy closing to random guessing.
    2) \tul{Imperceptibility}: the unlearnable examples shall be indistinguishable from the pristine clean examples, namely, the appended perturbations shall be imperceptible.
    3) \tul{Transferability}: the generated unlearnable example should be versatile enough to handle diverse scenarios, such as different facial datasets \cite{Yi2014webface, Liu2014CelebA, Cao2017VGGFace2} and different model architectures \cite{He2015ResNet, Howard2017MobileNets, Szegedy2015RethinkingTI} used by the attacker, even in a black-box scenario.
    4) \tul{Robustness}: since these facial images will be shared on social platforms, we need to account for the distortions caused by the transmission (\eg, JPEG compression and blurring). Moreover, the attacker may deliberately use adversarial training \cite{Goodfellow2014adversarial2} to undermine the effectiveness of unlearnable perturbations.
    5) \tul{Efficiency}: the generation speed is crucial for practical use, \eg, online processing requires fast generation. 
    However, as shown in \Tref{tab: Comparison}, existing methods of unlearnable examples \cite{huang2021unlearnable, fu2022robust, ren2023transferable} can not satisfy all the above demands, making them inapplicable
     for facial privacy protection in practice.

    To remedy the limitations of current methods, we propose \textbf{Segue}, a side-information guided generative unlearnable examples method. Specifically, we adopt an auto-encoder model to generate perturbations, which is more efficient than iterative gradient optimization used by previous approaches. The trained generator can generate perturbations for various scenarios without retraining, namely, can be once-trained multiple-used. Besides, we leverage side information to guide the generation process and improve transferability. The side information can be adapted to different protection scenarios based on prior knowledge. For instance, we use true labels of the to-be-protected category as side information if we can access it. Otherwise, we leverage K-means clustering \cite{Selim1984KMeans} on an unlabeled large facial dataset to get pseudo labels as side information. The side information helps to distinguish the target face from other faces, which is consistent regardless of the attacker’s training datasets and model architectures adopted. To enhance robustness against the transmission process, we further append a distortion layer to our training pipeline, wherein the distortion layer simulates possible channel losses and potential attacks in reality, such as JPEG compression, blurring, and adversarial training.

    We conduct extensive experiments to show that \textbf{Segue} can successfully meet the five requirements mentioned above. Specifically, our method induces a larger performance degradation of the attacker's model on clean examples, \eg, only 11.50\% accuracy on VGGFace10 while the best result for the other methods is 20.50\%. Furthermore, we compare the transferability across 5 different model architectures and 5 different facial datasets, the proposed method achieves a superior performance in most cases. For robustness, \textbf{Segue} performs better compared with the robust unlearnable example (RUE) \cite{fu2022robust} against adversarial training and other pre-processings. In terms of efficiency, \textbf{Segue} and non-trainable shortcuts (LSP) \cite{Yu2021AvailabilityAC} are much faster than other methods (1000$\times$). Finally, some ablation studies are also conducted to verify our design. 

    To summarize, the main contributions of our method are described as follows:
    \begin{itemize}
        \item We conclude five requirements for unlearnable facial examples for facial privacy protection in the real world: effectiveness, imperceptibility, transferability, robustness, and efficiency. Besides, we survey current methods and find that they cannot meet all these requirements.
        \item We propose \textbf{Segue}, which uses a once-trained multiple-used generative model to efficiently generate unlearnable examples. Side information and a distortion layer are also introduced to improve transferability and robustness.  
        \item Extensive experiments demonstrate that our approach surpasses current methods, especially in terms of  transferability, robustness, and efficiency.
    \end{itemize}
\section{Related Work}
\subsection{Facial Privacy Protection}
    Facial privacy protection aims to prevent unauthorized disclosure or use of individuals’ facial data, such as by face recognition technology \cite{Jiang2016Face1, Ranjan2016face2}, which can identify individuals without their consent. Existing protection methods against face recognition can be classified into two categories based on the stage of protection: the testing-stage protection and the training-stage protection. Testing-stage protection methods \cite{LowKey, Shan2020Fawkes, RN915} apply adversarial perturbations to images in the inference stage, making the model misclassify the perturbed images. However, these methods cannot stop the unauthorized usage of private data and cannot protect clean test images. On the other hand, training-stage protection methods \cite{Yu2021AvailabilityAC, Feng2019LearningTC, huang2021unlearnable} add perturbations to the training images, which aim to degrade the model’s performance on clean test images.
    In this paper, we focus on the latter protection strategy, which involves unlearnable examples \cite{huang2021unlearnable, ren2023transferable, fu2022robust} to interfere with the unauthorized training.

\subsection{Unlearnable Examples}
    Adversarial examples \cite{Szegedy2013adversarial1, Goodfellow2014adversarial2} are generated by a min-max optimization. Following this idea, Huang \etal~\cite{huang2021unlearnable} propose to use a min-min optimization strategy to generate unlearnable examples (UE). However, this strategy is not robust to adversarial training and needs the gradient information of the target model, which is not accessible in a black-box setting. To address the robustness issue, Fu \etal~\cite{fu2022robust} propose the robust unlearnable example (RUE), which integrates the adversarial training into the original generation process of perturbations by a min-min-max optimization strategy. However, this strategy is computationally expensive. Yu \etal~\cite{Yu2021AvailabilityAC} use several patches to synthesize linearly separable perturbations (LSP) without training, which is a strict black-box scenario. In \cite{ren2023transferable}, they propose transferable unlearnable examples (TUE) which focus on transferability rather than efficiency and robustness. They reduce the intra-class distance and increase the inter-class distance of perturbations to make them easily separable by even a linear classifier when they are added to any dataset.
    However, as shown in \Tref{tab: Comparison}, none of the current methods can satisfy all the five requirements for real-world applications simultaneously. Nevertheless, we take them as the baseline methods for subsequent comparison.

\section{Preliminary}
    \begin{figure*}[t]
    \centering
    \includegraphics[width=1.00\textwidth]{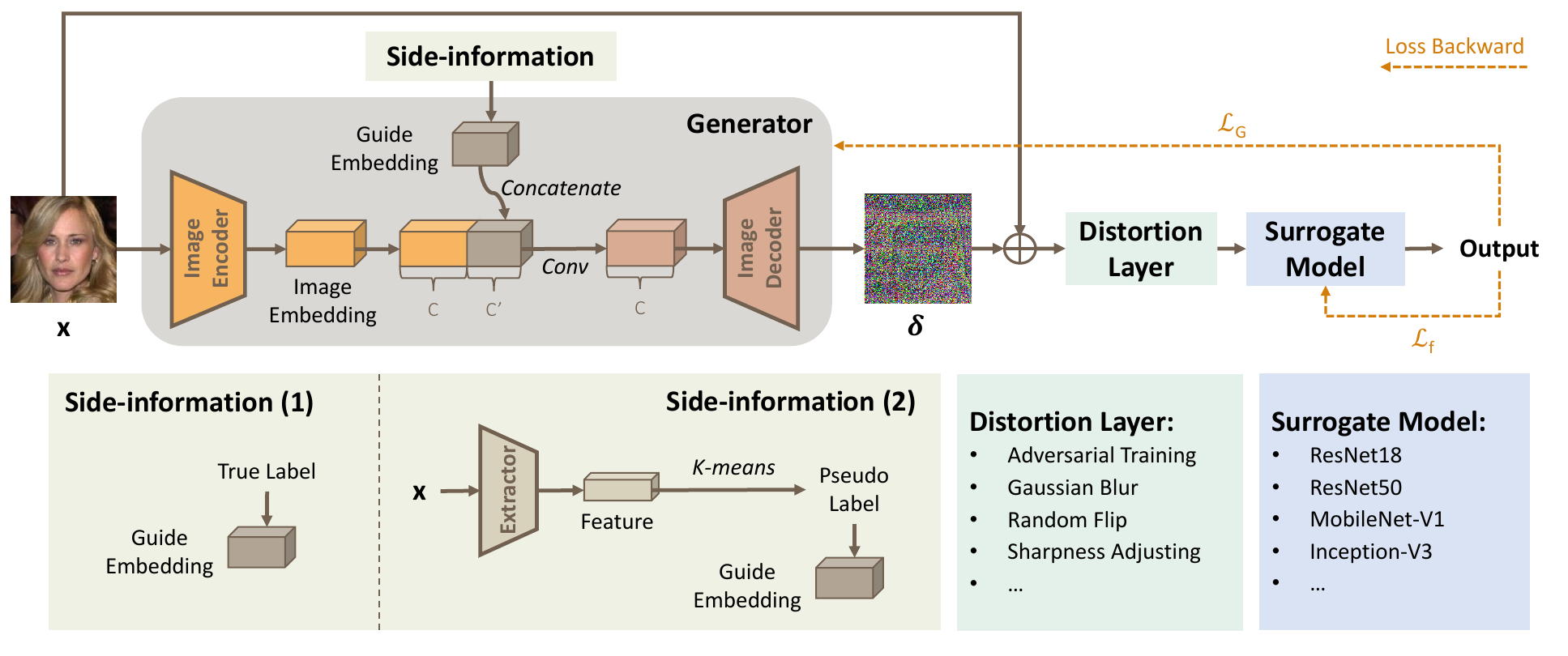}
    \vspace{-2em}
    \caption{
        The overall framework of Segue. 
        The generator consists of an encoder and a decoder. The image and side information are fused in the deep feature space. The side information could be the true label in supervised scenarios or the pseudo label in unsupervised scenarios.
        The distortion layer contains various processing to enhance the robustness of $\delta$.
        We train the generator and the surrogate model alternately.
    }
    \label{fig: pipeline}
    \vspace{-1em}
    \end{figure*}

\subsection{Formalized Description}
\subsubsection{Face Recognition}
    Face recognition is a type of image classification with DNNs. Suppose we have a clean dataset $\mathcal{D}=\{(x_i, y_i)\}^n_{i=1}$, which can be divided into a training set $\mathcal{D}_{train}$ and a testing set $\mathcal{D}_{test}$. For face recognition tasks, we usually train a neural network $f$ to fit the distribution of $\mathcal{D}_{train}$. The optimization can be described as follows:
    \begin{equation}
    \label{eq: Face Recognition}
        \underset{f}{\arg \min } \mathbb{E}_{(x,y) \sim \mathcal{D}_{train}}\left[\mathcal{L}\left(f(x), y\right)\right],
    \end{equation}
    where $\mathcal{L}$ can be the Cross-Entropy loss. After training, $f$ can be used to predict the label $y$ of sample $x$ in $\mathcal{D}_{test}$ since $\mathcal{D}_{test}$ has the same distribution as $\mathcal{D}_{train}$. 
    
\subsubsection{Unlearnable Examples}
    Huang \etal~\cite{huang2021unlearnable} propose a bi-level objective to generate perturbations to prevent the FR model from learning anything from the training data. They use the following objective:
    \begin{equation}
    \label{eq: Unlearnable Examples}
        \underset{f}{\arg \min } \mathbb{E}_{(x,y) \sim \mathcal{D}_{train}}\left[\underset{\delta}{\min } \mathcal{L}\left(f(x+\delta), y\right)\right],
    \end{equation}
    where the modified image $x+\delta$ is called as the unlearnable example and the set of all such images is the unlearnable dataset. $f$ acts as a surrogate model for the target model. The perturbation $\delta$ is bounded by $\|G(x)\|_{p} \leq \epsilon$ to guarantee that it is imperceptible to human eyes.
    They update $\delta$ and $f$ with an alternating training strategy. In each epoch, $f$ is trained on the perturbed data for a few steps to reduce the loss. Then $\delta$ is optimized to further lower the loss. The optimization stops when the loss on the perturbed data reaches a threshold, which means that there is no gradient for the target model to update its parameters.

\subsection{Threat Model}
\subsubsection{The Capability and Objective of the Attacker}
    Following current methods, we assume that
    the attacker wants to train a face recognition model as in \Eref{eq: Face Recognition} but only has access to the unlearnable dataset. As a result, the attacker trains the FR model with the unlearnable dataset instead of $\mathcal{D}_{train}$. To boost the model performance, the attacker may apply data augmentation techniques, such as Cutout, Mixup, and CutMix. Besides, the attacker may also use adversarial training, which can eliminate the non-robust features (\ie, the appended perturbation as shortcut features) from the input and make the model learn only the robust features.

\subsubsection{The Capability and Objective of the Defender}
    We consider a black-box scenario, where the defender has no knowledge of the target model, including parameters and architectures, used by the attacker. Instead, we use a surrogate model $f$ as an approximation. We can use the dataset labels as side information if we have them. However, the dataset labels are not necessary. We only need to know the number of identities $K$ in the dataset, which is used to cluster the image features and obtain the pseudo labels. We describe this process in more detail in the next section.
    
    Our goal is to protect the user's facial privacy. To do this, we optimize $\delta$ following \Eref{eq: Unlearnable Examples} and add $\delta$ to $\mathcal{D}_{train}$ to prevent the attacker from learning useful information from it. As a result, the FR model trained with $\delta+\mathcal{D}_{train}$ fails to recognize the images in $\mathcal{D}_{test}$ since the distributions have changed between them.

\subsection{Linear Separability}
    Linear separability means the samples can be easily separated by simple linear models. Yu \etal~\cite{Yu2021AvailabilityAC} show that the perturbations of the training-stage availability attacks \cite{huang2021unlearnable, Feng2019LearningTC} are all linearly separable. It is easier for models to learn the linear separability noise while ignoring the image information.
    
    We can achieve strong linear separability by adding class-wise perturbation to the samples. For example, we can add $\delta_i$ to all samples from the class $i$. Huang \etal~\cite{huang2021unlearnable} also points out that random class-wise perturbations can prevent the model from learning useful information. Random perturbation does not contain any information about the dataset or model and can be applied to any dataset or model, which implies linear separability ensures the transferability of perturbation. Therefore, we could increase the linear separability by enhancing the transferability of perturbation.

\section{Method}
\subsection{Overview}
    This section is organized as follows. First, we demonstrate what is side information and how to leverage it. Additionally, we describe our design of the once-trained multiple-used generator and the distortion layer. Finally, a two-stage training strategy is introduced.

\subsection{The Design of Side Information}
\label{sec: side-information}
   As mentioned above, we should increase the linear separability of perturbation to enhance its transferability. A simple and direct way is to use the dataset's true labels. Besides, we also address unlabeled facial data in the wild. We call them supervised and unsupervised scenarios, respectively.

    \begin{algorithm}[t]
    \caption{Two-stage Training Strategy}
    \label{alg}
    \KwIn{image $x$, side information $\hat{y}$, dataset $D$, perturbation boundary $\epsilon$, distortion layer $T$, generator $G$, surrogate model $f$, optimization steps $maxiter$, learning rate $\alpha_{f}$ and $\alpha_{G}$, epoch $E$ }
    \KwOut{generator $G$}
        \For{$epoch\leftarrow1$ \KwTo $E$}{
            \uIf{$epoch\%5 = 1$}{
                \For{$i\leftarrow1$ \KwTo $maxiter$}{
                    
                    $(x_i,y_i)\sim \mathcal{D}$\\
                    $\delta_i \leftarrow Clip(G(x_i),-\epsilon,\epsilon)$\\
                    $x_i' \leftarrow  x_i + \delta_i$\\
                    $\theta_{f,i+1}\leftarrow\theta_{f,i}-\alpha_{f}\nabla_{\theta_{f,i}}
                        \mathcal{L}_f(f(x_i'),\hat{y_i})$\\  
                }
            }
            \Else
            {
                \For{$i\leftarrow1$ \KwTo $maxiter$}{
                    
                    $(x_i,y_i)\sim \mathcal{D}$\\
                    $\delta \leftarrow Clip(G(x_i),-\epsilon,\epsilon)$\\
                    $x_i' \leftarrow  x_i + \delta_i$\\
                    $\theta_{G,i+1}\leftarrow\theta_{G,i}-\alpha_{G}\nabla_{\theta_{G,i}}
                        \mathcal{L}_G(f(x_i'),\hat{y_i})$\\  
                }
            }
        }
    \end{algorithm} 

\subsubsection{Supervised Scenario}
    In the supervised scenario, we use the dataset labels as side information. Inspired by \cite{Han2019MAN}, we concatenate the label embedding with the image embedding along the channel in the high-level feature space to guide the generation process (see \Fref{fig: pipeline}). Then, we use an extra convolution layer to reduce the channel dimension $C+C'$ back to $C$.
Specifically,
    we use a 16-bit binary vector to encode the label embedding. For instance, the label embedding is 0…0101 (13 zeros before 101) when $y$ is 5. 
    With a 16-bit label embedding, we can support up to $2^{16}$ identities in the dataset. Thus, we can handle most facial datasets \cite{Yi2014webface, Liu2014CelebA, Cao2017VGGFace2}. Moreover, the label embedding length is flexible. We can adjust it to any length depending on the number of classes in the dataset.

\subsubsection{Unsupervised Scenario}
    To address the challenge of obtaining labels in this scenario, we propose using pseudo labels generated by an unsupervised approach \cite{zhang2023Clusters}.
    First, we use an extractor trained on a large-scale facial dataset CelebA \cite{Liu2014CelebA} to extract facial features. Then we apply the K-means clustering method \cite{Selim1984KMeans} to cluster the facial features into $K$ groups and assign pseudo labels to them. Once we get the pseudo labels, we can concatenate them with image features as the supervised scenario. This frees us from manual labeling of the images and all we need to know is the number of classes $K$ in the dataset. Moreover, our method is robust to the accuracy of the k-means clustering method. As long as it is more than 80\%, there will be no impact on the transferability of perturbations across different datasets and models.

    \begin{figure*}[t]
    \centering
        \includegraphics[width=0.98\linewidth]{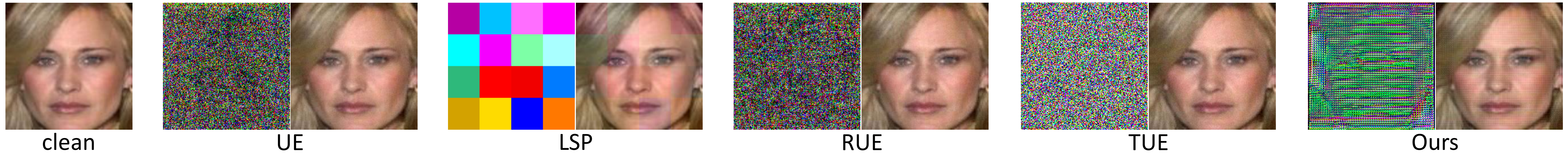}
    \caption{
    Visualization of different unlearnable examples and the corresponding residual compared with the clean image.
    }
    \label{fig: iqa}
    \end{figure*}

    \vspace{-0.6em}
\subsection{Once-trained Multiple-used Generator}
    As shown in \Fref{fig: pipeline}, the generator  $G$ encodes the input image into the image embedding and decodes it into a perturbation $\delta$. We use convolutional layers with kernel size (3$\times$3), batch normalization, and ReLU activation for both the encoder and the decoder. The image embedding is fused with the guide embedding in the high-level feature space to guide the generation of $\delta$. 
%
    Unlike previous methods \cite{huang2021unlearnable, ren2023transferable} that optimize the perturbation directly, we optimize a generator that produces perturbations according to inputs. This design allows us to generate perturbations for different datasets with different numbers of classes using one generator. For example, we can generate perturbation for WebFace50 (a dataset with 50 identities) with the generator trained on WebFace10 (a dataset with 10 identities). In contrast, most existing methods need to retrain the perturbations for different datasets. In other words, this design is more efficient.

\subsection{Distortion Layer}
\label{sec: Distortion Layer}
    To enhance the robustness of unlearnable perturbations against possible distortions in transmission, we use a distortion layer during training.
    RUE \cite{fu2022robust} adopts a min-min-max framework which introduces adversarial training to increase the perturbation generation difficulty.
    Actually, adversarial training can be seen as a form of data augmentation. Thus, we use a distortion layer to augment the data with common transformations including adversarial training, Gaussian blurring, random flip, etc. (see \Fref{fig: pipeline}). The distortion layer perturbs the data in the high-dimensional space, forcing the generator $G$ to find a more robust perturbation, which makes all points near the perturbation become unlearnable in the high-dimensional space.
    
\subsection{Two-stage Training Strategy}
\label{sec: loss}
    As shown in \Aref{alg}, we alternately train the surrogate model $f$ and the generator $G$, where ResNet18 is adopted as the default surrogate model $f$. In the first stage, we train $f$ for $maxiter$ (iterations over the entire dataset) constrained by $\mathcal{L}_f$, which encourages the perturbed image $x+G(x)$ to be classified correctly (as $\hat{y}$) by $f$:
    \begin{equation} 
    \label{eq: l_f}
         \mathcal{L}_f=CE(f(x+G(x)),\hat{y}),
    \end{equation}  
        where $\hat{y}$ denotes the side information, including true-label and pseudo-label, and $CE$ denotes Cross-Entropy loss.
    In the second stage, we update $G$ for four epochs. 
    The loss function for  $G$ consists of two terms, namely, $\mathcal{L}_{G1}$ tries to reduce the loss of $f$ on unlearnable examples, while $\mathcal{L}_{G2}$ aims to constrain the magnitude of the perturbation:
    \begin{gather} 
    \label{eq: l_G}
         \mathcal{L}_G=\alpha\cdot \mathcal{L}_{G1} + \beta\cdot \mathcal{L}_{G2},\\
         \mathcal{L}_{G1}=CE(f(x+G(x)),\hat{y}),\\
         \mathcal{L}_{G2}=\mathbb{E}_{x} (\|G(x)\|_{2}),
    \end{gather}
where $\alpha$ and $\beta$ control the relative importance of each objective.  We stop the whole training after 20 epochs or when the loss of $f$ on the unlearnable dataset is below 0.001.

\section{Experiments}
\label{sec: exp}
    We evaluate the proposed method \textbf{Segue} on various aspects, including effectiveness, imperceptibility, transferability, robustness, and efficiency. We compare \textbf{Segue} with current methods and show its advantages. We also conduct ablation studies to validate our design choices.
    
\subsection{Experimental Settings}

    \begin{table}[t]
        \small
    \centering
    \setlength{\tabcolsep}{1mm}{
    \begin{tabular}{ccccc}
    \toprule
    Sub-dataset& \# IDs&Source dataset&$\mathcal{D}_{train}$&$\mathcal{D}_{test}$\\
    \hline
    WebFace10 & 10 & WebFace  & 1300 & 200\\
    WebFace10 $\dagger$ & 10 & WebFace  & 1300 & 200\\
    WebFace50 & 50 & WebFace  & 6500 & 1300\\
    VGGFace10 & 10 & VGGFace2 & 1300 & 200\\
    CelebA10  & 10 & CelebA   & 200  & 100\\
    \bottomrule
    \end{tabular}
    }
    \caption{The details of datasets.}    
    \label{tab: subdataset1}
    \end{table} 
    
\subsubsection{Datasets}
\label{sec: datasets}
    We use three face image datasets: WebFace \cite{Yi2014webface}, VGGFace2 \cite{Cao2017VGGFace2}, and CelebA \cite{Liu2014CelebA}. For ease of implementation on each dataset, we randomly select some categories to construct the final sub-datasets and resize all images to 224$\times$224.  More details can be found in \Tref{tab: subdataset1}.
    

\subsubsection{Metric}
    We adopt \textit{clean test accuracy}, \ie, the performance of the attacker's model on clean datasets, where the model is trained on unlearnable datasets. The lower the test accuracy, the more effective unlearnable examples are. If the accuracy is close to $\frac{100\%}{\#IDs}$, the model learns nothing from the unlearnable dataset, just like random guessing. 

\subsubsection{The Baselines}
    We compare our method with three gradient-based methods: UE \cite{huang2021unlearnable}, RUE \cite{fu2022robust}, and TUE \cite{ren2023transferable}. Besides, the model-agnostic method LSP \cite{Yu2021AvailabilityAC} is also considered for comparison. We follow their official code to reproduce them.
    
\subsubsection{Implementation Details}
    We limit the perturbation to $\|\delta\|_{\infty} \leq \epsilon = 8/255$, which is imperceptible to humans. We use Adam optimizer with an initial learning rate of 0.0005 for both the surrogate model $\alpha_{f}$ and the generator $\alpha_{G}$. We set $\alpha$ and $\beta$ in \Eref{eq: l_G} to 1 and 0.001, respectively. We highlight the best results in \textbf{bold}. Unless specified, we use ResNet18 and WebFace10 by default, which can be seen as a white-box setting. To evaluate transferability across different models and datasets, we conduct control experiments.

    We update the surrogate model $f$ for one epoch and then update the generator $G$ for five epochs and repeat. $maxiter$ represents the number of all samples divided by the batch size.
    The distortion layer consists of adversarial training, Gaussian blur, random horizontal flip, random vertical flip, and sharpness adjusting. For adversarial training, we set the $\rho_{d}=1/255$ as default and adjust it from $\rho_{d}=0/255$ to $\rho_{d}=4/255$ in the experiment of robustness against adversarial training. For Gaussian blur, we set kernel size to (3,3) and sigma to 0.2. For sharpness adjusting, we set the sharpness factor to 2. For random horizontal flip and random vertical flip, we set the probability of the image being flipped to 0.1.
    We resize the images to 32$\times$32 on CIFAR10\cite{Krizhevsky2009cifar} and 224$\times$224 on other sub-datasets.

    \begin{table}[t]
    \setlength{\tabcolsep}{2.4mm}{
    \begin{tabular}{ccccc}
    \toprule
    Methods&WebFace10&WebFace50&VGGFace10\\
    \hline
    CLEAN& 75.00 & 80.60 & 83.00\\
    UE \cite{huang2021unlearnable}&  12.50 &  3.50 & 20.50\\
    LSP \cite{Yu2021AvailabilityAC}&31.50&9.30&57.50 \\
    RUE \cite{fu2022robust}&  11.50 &  7.40 & 30.00\\
    TUE \cite{ren2023transferable}&  33.50 & 11.20 & 82.00 \\
    Ours&\textbf{10.50}&\textbf{2.50}&\textbf{11.50}\\
    \bottomrule
    \end{tabular}
    }
    \caption{Comparison of effectiveness (clean test acc \%  $\downarrow$) among different methods. Experiments are conducted with ResNet18 on three different facial datasets.}
    \label{tab: acc}
    \end{table} 

    \begin{table}[t]
    \setlength{\tabcolsep}{2.9mm}{
    \begin{tabular}{ccc}
    \toprule
    Methods&CIFAR10&ImageNet10\\
    \hline
    CLEAN& 91.67 & 71.00 \\
    UE \citep{huang2021unlearnable}& 19.93 & 30.00\\
    LSP (patchsize=8) \citep{Yu2021AvailabilityAC}& 17.07 & 64.00 \\
    LSP (patchsize=56) \citep{Yu2021AvailabilityAC}& \diagbox[]{}{} & 28.50 \\
    RUE \citep{fu2022robust}& 15.18 & 24.50\\
    TUE \citep{ren2023transferable}& 11.25 & 60.50 \\
    Ours& \textbf{10.12} & \textbf{14.00} \\
    \bottomrule
    \end{tabular}
    }
    \caption{Comparison on CIFAR10 and ImageNet10.}
    \label{tab:acc—noface}
    \end{table} 

    \begin{table*}[t]
    \setlength{\tabcolsep}{6.0mm}{
    \begin{tabular}{c c c c c c}
    \toprule
    Methods & ResNet18 & ResNet50 & MobileNet-V1 & Inception-V3 & EfficientNet-b1 \\ \hline
    UE \cite{huang2021unlearnable}    & 14.50     & 14.50     & 15.50         & 73.00          & 28.00              \\ 
    LSP  \cite{Yu2021AvailabilityAC}  & 31.50     & 32.50     & 18.50         & 56.00           & 52.50            \\ 
    RUE  \cite{fu2022robust}  &  19.00      & 27.50     &  17.00       &   77.00         &    27.00      \\ 
    TUE  \cite{ren2023transferable}   & 33.50     & 70.00       & 15.50         & 69.00           & 67.50            \\ 
    Ours   & \textbf{10.50}     & \textbf{12.50}     & \textbf{11.00}           & \textbf{10.50}         & \textbf{12.00}              \\ 
    \bottomrule
    \end{tabular}
    }
    \caption{Transferability for different models (clean test acc \% $\downarrow$). Defenders use ResNet18 and attackers use five models.
    }
    \label{tab: tran-model}
    \end{table*} 

    \begin{table*}[t]
    \setlength{\tabcolsep}{6.5mm}{
    \begin{tabular}{cccccc}
        \toprule
        Methods&WebFace10&WebFace10$\dagger$&WebFace50&VGGFace10&CelebA10\\
        \hline
        UE \cite{huang2021unlearnable} & 12.50  & 14.50 & $\backslash$ & 21.50 & 44.00 \\
        LSP \cite{Yu2021AvailabilityAC}& 31.50  & 35.00 & \textbf{9.30}& 57.50 & 74.00 \\
        RUE  \cite{fu2022robust}      & 17.00  & 26.50  & $\backslash$ & 78.50 & 78.50 \\ 
        TUE \cite{ren2023transferable}  & 33.50 & 52.00 & $\backslash$ & 53.00 & 59.00 \\
        Ours&\textbf{10.50}&\textbf{11.50}& 13.50 &\textbf{13.00}&\textbf{17.00}\\
        \bottomrule
    \end{tabular}
    }
    \caption{Transferability across different datasets (clean test acc \% $\downarrow$). Perturbations are trained on WebFace10 and then added to the other different datasets. WebFace10$\dagger$ owns 10 non-overlapped categories with WebFace10.
    }
    \label{tab: tran-dataset}
    \end{table*} 

    \begin{table}[t]
    \setlength{\tabcolsep}{3.8mm}{
    \begin{tabular}{cccc}
        \toprule     Methods&PSNR($\uparrow$)&SSIM($\uparrow$)&LPIPS($\downarrow$)\\
        \hline
        UE \cite{huang2021unlearnable}  & 32.37 & 0.754 & 0.205\\
        LSP \cite{Yu2021AvailabilityAC} & 31.53 & \textbf{0.968} &  \textbf{0.049}\\
        RUE \cite{fu2022robust} & \textbf{32.45} & 0.763 & 0.188\\
        TUE \cite{ren2023transferable} & 30.18 & 0.651 & 0.310\\
        Ours& 30.54 & 0.673 & 0.159\\
        \bottomrule
    \end{tabular}
    }
    \caption{Comparison of imperceptibility on WebFace10.} 
    \label{tab: iqa}
    \end{table} 
    
\subsection{Effectiveness}
    \Tref{tab: acc} shows that our method achieves the best results among three facial datasets, which successfully reduces the accuracy closing to $\frac{100\%}{\#IDs}$. 
    We test different sizes of patches for LSP and show the best result with the size equal to 56$\times$56.
%
    We explain why the results vary across datasets as follows: WebFace50 has more categories, so the difficulty of learning for the classifier goes up and the accuracy is lower. VGGFace10 has higher image quality, so the classifier learns the face features more easily and the accuracy is higher. 
    As shown in \Tref{tab:acc—noface}, \textbf{Segue} also achieves superior performance on non-facial datasets such as CIFAR10 and ImageNet10.

\subsection{Imperceptibility}
    In \Tref{tab: iqa}, we use three metrics to measure image quality: PSNR, SSIM, and LPIPS.
    \Fref{fig: iqa} also provides some visual examples, where perturbations are magnified to 30$\times$.
%
    UE, TUE, and RUE optimize the perturbations from random noises. LSP \cite{Yu2021AvailabilityAC} composes the perturbation with several patches, but they are visible to the human eye. We use an encoder-decoder structure generator to create perturbations that preserve distinct facial features. Therefore, our perturbations are more diverse and relevant to each sample than other methods. Overall, we achieve comparable imperceptibility compared with other methods.

    \begin{table*}[t]
    \small
    \setlength{\tabcolsep}{2.8mm}{ 
    \begin{tabular}{ccccccccccccc}
        \toprule
        Adv. Train.& Clean& UE & \multicolumn{5}{c}{RUE }& \multicolumn{5}{c}{Ours} \\
        \cmidrule(r){4-8} \cmidrule(r){9-13}
        $\rho_{a}$& & 
        &$\rho_{d}$=0&1/255&2/255&3/255&4/255
        &$\rho_{d}$=0&1/255&2/255&3/255&4/255\\
        \hline
        0    & 75.00&12.50&11.50&13.50&13.00&12.50&14.50
        &\textbf{10.50}&12.50&12.50&12.50&13.50\\
        
        1/255& 68.00&18.00&17.50&17.00&15.50&18.00&17.50
        &13.50&13.50&11.50&\textbf{11.00}&12.50\\
        
        2/255& 65.00&69.00&26.00&24.50&19.50&23.50&22.50
        &15.50&14.50&15.00&13.50&\textbf{12.50}\\
        
        3/255& 63.50&74.50&69.50&68.50&61.50&53.50&58.00
        &29.00&27.00&16.00&16.50&\textbf{14.50}\\
        
        4/255& 65.50&69.50&71.00&66.50&62.00&63.50&63.00
        &34.00&35.50&21.00&19.50&\textbf{16.50}\\
        \bottomrule
    \end{tabular}
    }
    \caption{Robustness against adversarial training (clean test acc \% $\downarrow$). The perturbation budget used in adversarial training by the attacker is $\bm{\rho_{a}}$, while the perturbation budget used by the defender is $\bm{\rho_{d}}$.}
    \label{tab: adv}
    \end{table*} 

\subsection{Transferability}
\subsubsection{Different Models}
\label{sec: tran-models}
    All methods generate unlearnable examples based on the surrogate model ResNet18, and the attacker can adopt different model architectures for training. As show in \Tref{tab: tran-model}, we test on five architectures: ResNet18, ResNet50 \cite{He2015ResNet}, MobileNet-V1 \cite{Howard2017MobileNets}, Inception-V3 \cite{Szegedy2015RethinkingTI}, and EfficientNet-b1 \cite{Tan2019EfficientNetRM}, and our method performs well in all cases. For deeper networks like Inception-V3, other methods fail for privacy protection. We explain that different convolutional kernel sizes of Inception-V3 may filter their perturbations and capture rich image features.


\subsubsection{Different Datasets}
\label{sec: tran-dataset}
    Similarly, in \Tref{tab: tran-dataset}, we generate perturbations based on WebFace10 and conduct evaluations on different datasets.
    CelebA10 has fewer samples in each class, which requires higher linear separability of the perturbations, but our method still lowers the accuracy to 17\%. TUE and UE cannot transfer the perturbations to WebFace50, because they must fix the shape of the perturbation before optimizing. Therefore, they can only transfer to smaller datasets, which limits their applicability.

    \begin{figure*}[t]
    \includegraphics[width=0.99\textwidth]{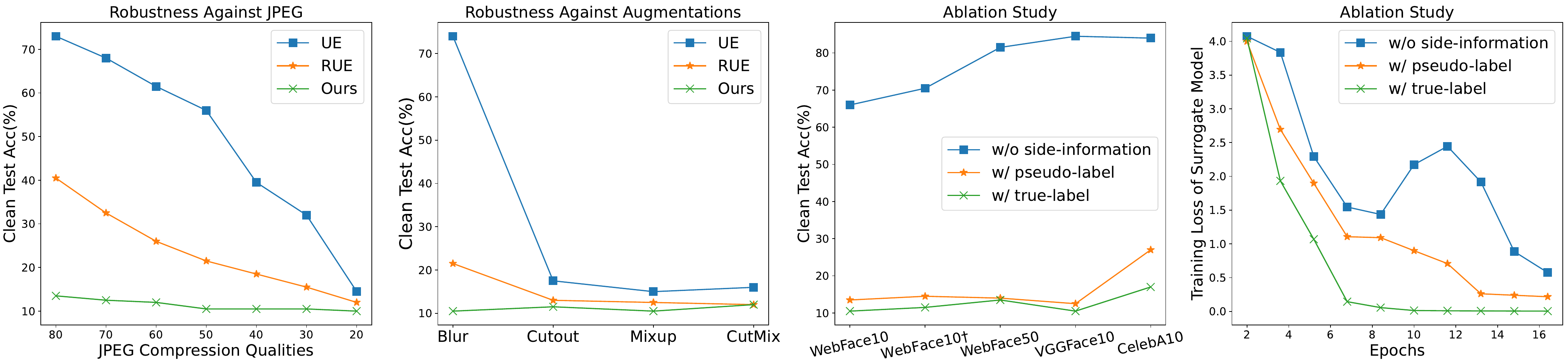}
    \caption{From left to right: (1) Robustness against JPEG compression. A lower quality indicates a higher compression ratio. (2) Robustness against different data augmentations. (3) The influence of side information on transferability. (4) The influence of side information on the training loss of the surrogate model. For all results, the lower the numerical value, the better. }
    \label{fig: all}
    \end{figure*}

\subsection{Robustness}
\subsubsection{Adversarial Training}
    Adversarial training can effectively remove the non-robust noise from the input \cite{Goodfellow2014adversarial2}.
    The attacker employs adversarial training with $\rho_{a}$ to remove the perturbations from the images, while we use $\rho_{d}$ in the distortion layer to improve the robustness of perturbations. When $\rho_{a}$ and $\rho_{d}$ are 0, it means that neither the attacker nor the defender uses adversarial training. \Tref{tab: adv} shows that our method can still achieve good effectiveness with 16.5\% clean data accuracy, even if the attacker uses adversarial training with $\rho_{a}=4/255$.
\subsubsection{JPEG Compression}
    JPEG compresses images by dividing them into 8×8 pixel blocks, transforming them into frequency components, and discarding some of the less important ones. A higher compression ratio requires more robust perturbations. 
    RUE uses $\rho_{d}=4/255$ as the adversarial perturbation radius. \Fref{fig: all} (1) shows that our method can maintain effectiveness against all quality settings, while other methods fail against low-quality settings.
    
\subsubsection{Data Augmentation}
    We implement different augmentations as follows:
     we use a kernel size of 5 and a standard deviation of 1.0 for Gaussian blurring. For Cutout \cite{Devries2017Cutout}, we use 2 patches with a length of 112, which is half of the image size 224. For Mixup \cite{Zhang2017mixup}, we randomly select a pair of images to mix up and $\lambda$ takes values from the beta distribution in the range of [0,1]. For CutMix \cite{Yun2019CutMix}, we apply Mixup on the Cutout region with the same setting in Cutout. \Fref{fig: all} (2) shows that our method is robust against all these data augmentations.

    \begin{table}[t]
    \setlength{\tabcolsep}{14mm}{
    \begin{tabular}{cc}
        \toprule
        Methods&Time ($s$)\\
        \hline
        UE \cite{huang2021unlearnable}& $\sim$2.1k\\
        LSP \cite{Yu2021AvailabilityAC}& 4.5\\
        RUE \cite{fu2022robust}& $\sim$6.7k\\
        TUE \cite{ren2023transferable}&  $\sim$7.4k\\
        Ours& \textbf{2.2}\\
        \bottomrule
    \end{tabular}
    }
    \caption{Comparison of efficiency on WebFace10.}
    \label{tab: time}
    \end{table} 

\subsection{Efficiency}
    We use a server with a single A6000 GPU and an Intel Xeon Gold 6130 CPU. Our methods only need one-step inference with a trained generator. LSP generates perturbations without training. UE runs 100 SGD updates for the outer problem and 20 SGD updates for each target example in the inner problem, as in \Eref{eq: Unlearnable Examples}. TUE trains the model parameters for 50 SGD updates and optimizes the perturbations for one SGD update by PGD-20 after every 1/4 update.
    As shown in \Tref{tab: time}, our method can generate unlearnable examples for the entire WebFace10 dataset much faster than other methods that rely on gradient optimization, achieving a speedup of over 1000$\times$.

\subsection{Ablation Study}
    \Fref{fig: all} (3) shows that side information improves the transferability of perturbations, and the true label is much better than the pseudo-label. 
    Besides, we also analyze the influence of side information on training convergence. As in \Fref{fig: all} (4), without side information, the generator’s optimization function is hard to converge and the surrogate model’s training loss fluctuates. We explain that side information acts as a prior to narrow the generator’s search space, which speeds up the training convergence and provides more precise guidance.

\section{Conclusion}
    In this paper, we present a novel method \textbf{Segue} for facial privacy protection with unlearnable examples, which satisfies five requirements: effectiveness, imperceptibility, transferability, robustness, and efficiency. 
    The proposed method uses generative models with side information to create unlearnable examples that are hard to recognize by face recognition models. We have shown that our method can transfer well across different datasets and models, and can resist various attacks and distortions. Moreover, our method can generate unlearnable examples much faster than most existing methods, achieving up to 1000$\times$ speedup. 
    We believe our work can provide a new perspective and a practical solution for facial privacy protection in the real world.

{
    \small
    \bibliographystyle{ieeenat_fullname}
    \bibliography{main}
}


\end{document}